\documentclass[amsmath,amssymb,aps,prb,twocolumn,superscriptaddress]{revtex4-2}

\usepackage{graphicx}
\usepackage[dvipsnames]{xcolor}
\usepackage{array}
\usepackage{float}

\begin{document}

\title{A dielectric microcylinder makes a nanocylindrical trap for atoms and ions}

\author{V.~Klimov}
\email[]{klimov256@gmail.com}
\affiliation{Department of Optics, Lebedev Physical Institute, Russian Academy of Science, 53, Leninski Prospekt, 119991, Moscow, Russia}
\author{R.~Heydarian}
\email[]{reza.heydarian@aalto.fi}

\author{C.~Simovski}
\affiliation{Department of Electronics and Nano-Engineering, Aalto University, P.O. Box 15500, FI-00076 Aalto, Finland}
\altaffiliation[Also at ]{Faculty of Physics and Engineering, ITMO University, 199034, Birzhevaya line 16, Saint-Petersburg, Russia}

\date{\today}

\begin{abstract}
In the diffraction of visible light by a dielectric microcylinder{,} packages of evanescent waves always arise. However, single plane wave incidence corresponds to rather small impact of evanescent waves outside the cylinder.  
In this paper, we theoretically show that a pair of plane waves impinging a glass microcylinder under certain conditions may correspond to much higher impact of the evanescent waves. Namely, the interference of the evanescent waves with the propagating ones results in the suppression of the electromagnetic field in an area with very small cross section. This area is located in free space at a substantial distance from the {rear side of the microcylinder and along its axis}. It may serve a linear optical trap for cold atoms and ions.

\end{abstract}

\maketitle

\section{Introduction}

\begin{figure*}[ht]
\centering

\includegraphics[width=15cm]{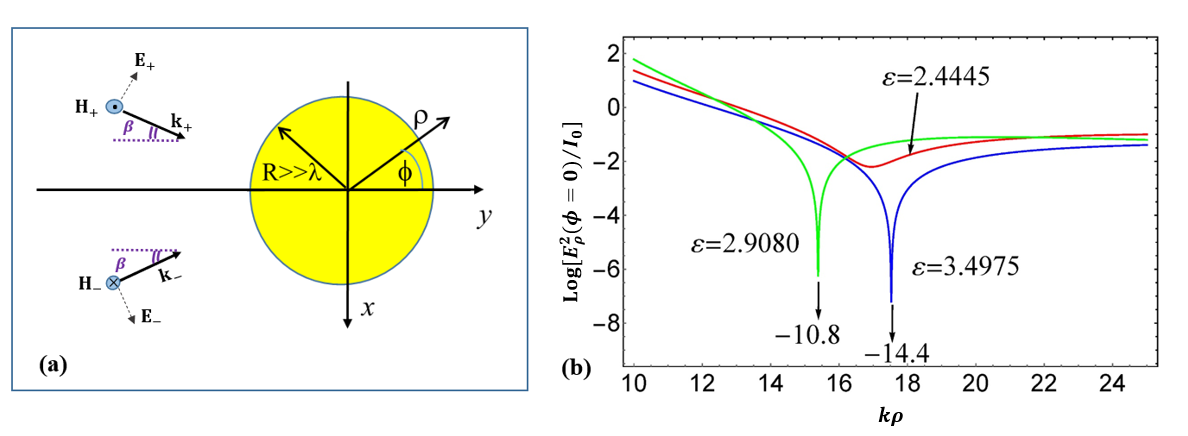}
\caption{(a) Two plane waves with the same electric field amplitude and wave vectors $\bf k_{\pm}$ of the same length impinge a dielectric microcylinder. (b) Normalized intensity $I/I_0$ in the log scale on the axis $y$ behind the cylinder versus normalized coordinate $k\rho\equiv k y$ for different values of the permittivity $\varepsilon$. Fixed parameters: $\beta=0.01$,
$kR=10$.}
\label{Pic1}
\end{figure*}

\begin{figure*}[ht]
\centering
\includegraphics[width=15cm]{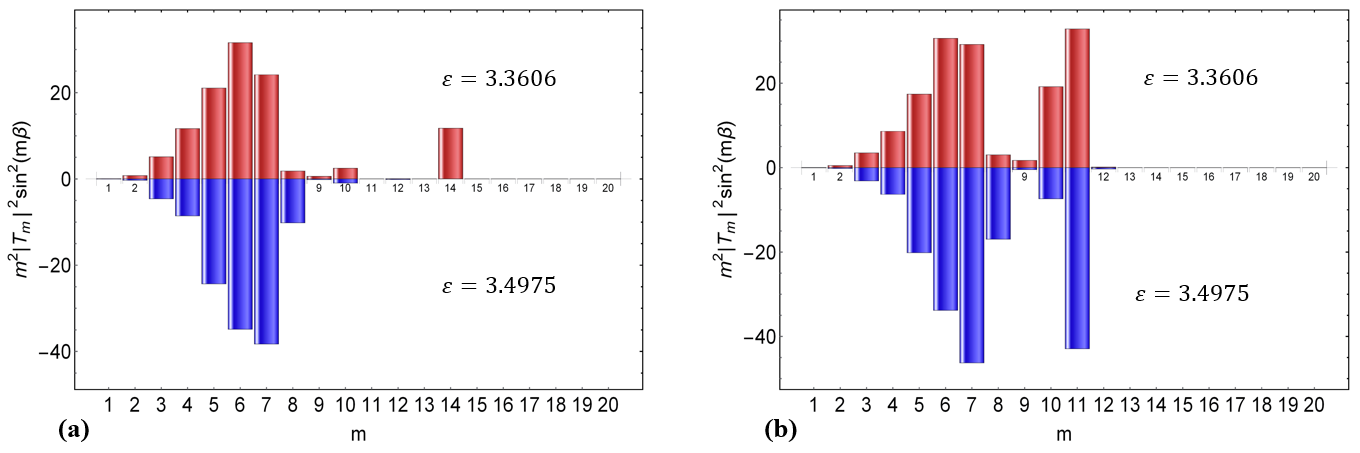}
\caption{Modal amplitudes for $kR=10$ when (a) $\beta=\pi/11$  and (b) $\beta=\pi/14$. }
\label{beta effect}
\end{figure*}

The most known near-field effect of the visible light diffraction by a dielectric microcylinder, is the waist of the called photonic nanojet 
\cite{PNJ}. This nanojet is a wave beam whose waist is centered at the rear edge of the particle \cite{PNJ,Itagi,Taflove,Taflove1}. In this area, the field spatial spectrum comprises a noticeable evanescent-wave component \cite{Itagi,Taflove1} that implies a nonzero longitudinal component of the electric field in the case of the TM-incidence \cite{Itagi, Taflove1}. However, the effective width of this waist for a cylinder is not very subwavelength (of the order of $0.5\lambda$) and the local enhancement of the electric intensity is modest ($3-4$). Such near-field effect can be called slightly subwavelength -- the contribution of evanescent waves into the field in the region of the waist is not dominant \cite{Taflove1}. It is dominant at the frequencies of well-known resonances -- those of whispering gallery modes and at Mie resonances of the microcylinder
\cite{Sergei, Kavungal, Astratov1}. However, the spatial regions where evanescent waves dominate at these resonances are located inside the cylinder
and this domination implies high values of the local electric intensity compared to that of the incident wave. Briefly, for a dielectric microparticle (cylinder or sphere) known near-field effects are usually effects of subwavelength field concentration \cite{Astratov1,Zhou,Klimov}. 

However, there are no theoretical restrictions for pronounced near-field {effects outside} a microsphere or a microcylinder. If a cylinder is made of a dielectric material with the refractive index, say, $n=\sqrt{\varepsilon}=1.5-1.8$ the spatial variation of the induced eigenmodes inside and outside the cylinder have the same scale. {Inside a microcylinder, the fields are} expressed via Bessel's functions $J_m(k_c\rho)$ (here $k_c=kn$ and $k$ is the wave number in free space) and outside -- via Hankel's functions $H^{(1)}_m(k\rho)$ which comprise the Bessel component as well. Though it is usually thought that all practically important near-field effects for a glass  microparticle (cylinder or sphere) are effects {of subwavelength} concentration of the electromagnetic fields, we will show that it is not so. We will report an amazing near-field effect which arises for a microcylinder and has nothing to do with the field concentration in it. It arises at the frequencies slightly shifted from those of high-order Mie resonances.

\section{Spatial Fano resonance behind the microcylinder}

Consider a microcylinder of radius $R\gg\lambda$ impinged by a wave beam consisting of two TM-polarized plane waves with the same magnitude of  magnetic fields but different phases: 
\begin{equation}
{\bf H}_{\pm}=\pm H_0e^{ik(\pm x\sin\beta+y\cos\beta)} {\bf a}_{z},
\end{equation}
as it is depicted in Fig.~\ref{Pic1}(a). Here we adopt the time dependence $\exp(-i\omega t)$. {The two incident waves are schematically shown in Fig. 1.} Let the wave vectors $\bf k_{\pm}$ be tilted to the axis $y$ with the sharp angle $\beta$ so that the resultant transversely bisects the cylinder. Expanding these two plane waves into cylindrical Bessel functions $J_m(k\rho)$ and uniting the terms of the same order we deduce after some algebra the following expression for the magnetic field of the incident beam (${\bf H}_{ib}$):
\begin{equation}
{\bf H}_{ib}=-4iH_0\sum\limits_{m=1}^{+\infty}i^{m}J_m(k\rho)\sin {m\beta}\sin {m\phi} {\bf a}_{z},
\label{eq1}
\end{equation}
The corresponding electric field of the incident beam can be easily obtained by differentiation of every series term in (\ref{eq1}):
\begin{equation}
{\bf E}_{ib}=\frac{i}{\omega\epsilon}\nabla\times {\bf H}_{ib}=\frac{i}{\omega\epsilon}[\frac{1}{\rho}\frac{\partial H_{ib}}{\partial \phi} {\bf a}_{\rho}-\frac{\partial H_{ib}}{\partial \rho} {\bf a}_{\phi}], 
\label{eq3}
\end{equation}
where ${\bf a}_{\rho}$ and ${\bf a}_{\phi}$ are two other unit vectors of the cylindrical coordinate system.

For the total magnetic field $\bf{H}$  outside the cylinder we further obtain
\begin{eqnarray}
{\bf H}=-4iH_0\sum\limits_{m=1}^{\infty} i^{m}[J_m(k\rho)+T_mH^{(1)}_m(k\rho)]\nonumber \\*
\times\sin {m\beta}\sin{m\phi} {\bf    a}_{z},
\label{eq4}
\end{eqnarray}
Total electric field has two components ${\bf E}=E_\rho {\bf a}_{\rho}+ E_\phi {\bf a}_{\phi}$, for which we have
\begin{eqnarray}
    E_\rho=4H_0\eta\sum\limits_{m=1}^{\infty} i^{m} \frac{m}{k\rho} [J_m(k\rho)+T_mH^{(1)}_m(k\rho)] \nonumber \\*
    \times\sin{m\beta}\cos{m\phi}\nonumber 
\end{eqnarray}
\begin{eqnarray}
    E_\phi=4H_0\eta\sum\limits_{m=1}^{\infty} i^{-m} \left[ J'_m(k\rho)+T_m \left(H^{(1)}_m(k\rho)\right)'\right] \nonumber \\*
    \times\sin{m\beta}\sin{m\phi},
    \label{eq5}
\end{eqnarray}
Here prime for the cylindrical functions means the derivative with respect to the argument 
$k\rho$ and the coefficients 
$$
T_m=\frac{k_cJ_m'(kR)J_m(k_cR)-kJ_m'(k_cR)J_m(kR)} {k J_m'(k_cR) H^{(1)}_m(kR)-k_c J_m(k_cR)[H^{(1)}_m(kR)]'}
$$
are the Mie coefficients. 
The factor {$\sin{m\beta}$} in every term of series (\ref{eq4} and \ref{eq5}) nullifies both 
$H_z$ and $E_{\phi}$ on the axis $y$ and drastically changes the total field spatial distribution compared to the single-wave incidence. Moreover, the choice of proper $\beta$ allows one to turn the selected $m-th$ mode on (when $\sin m\beta=\pm 1$) or off (when $\sin m\beta=0$). The relative contribution {of the $m-th$ mode into the intensity of scattered field on axis $y$ is shown in Fig. \ref{beta effect}. From Fig. \ref{beta effect} one can see that if the cylinder permittivity is chosen to be $\varepsilon=3.3606$, by changing $\beta$ from $\pi/14$ to $\pi/11$ the contribution of the TE$_{11,2}$ resonance mode of the cylinder will be nullified. Also if the cylinder permittivity is $\varepsilon=3.4975$, by changing $\beta$ from $\pi/11$ to $\pi/14$ the same will happen for TE$_{14,1}$ resonance mode (here and below the resonance mode is characterized by two numbers since there is no propagation along $z$).} In what follows we optimize $\beta$ so that to provide stronger impact of the selected resonant mode.

Near the high-order Mie resonance one of the series terms with number $m=M\gg 1$ describing the $M$-th mode (strictly speaking, quasi-mode) of the cylinder dominates over any other modes. In the case of a single-wave incidence, the almost-resonant $M$-th mode weakly interferes with the spatial quasi-continuum of non-resonant terms. Numerical analysis shows that in this case the sum of all non-resonant terms has the smaller magnitude than the magnitude of the almost-resonant mode. Outside the cylinder the evanescent part of the $M$-th mode rapidly decays, whereas its {propagating} part (cylindrical wave corresponding to the large-argument asymptotic of $H^{(1)}_M(k\rho)$) varies in sync with the quasi-continuum. On the contrary, when two slightly tilted plane waves illuminate the cylinder with opposite phases, and the  factor $\sin{m\beta}$ in each term of the Mie series arises the magnitude of the quasi-continuum of other modes $m\ne M$ has the magnitude of the same order as that of the $M$-th mode in the whole region of our interest -- near the back side of the illuminated cylinder.

\begin{figure}[h]
\centering
\includegraphics[width=8.5cm]{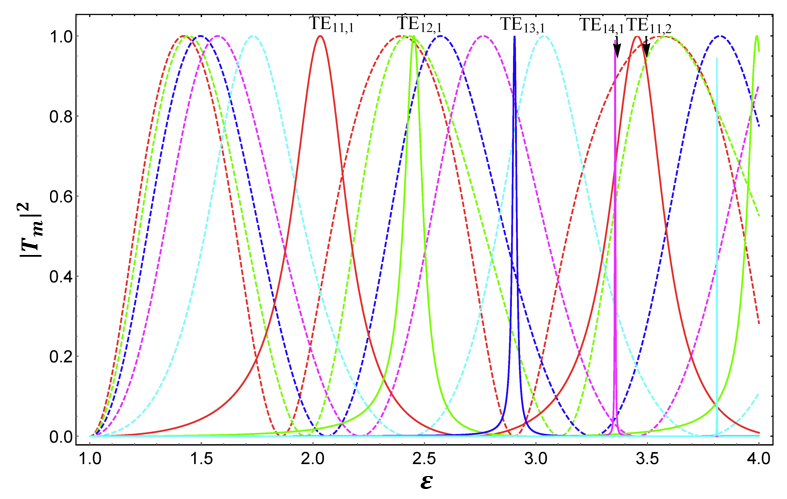}
\caption {The dependence of Mie coefficients $T_m$ for $m=6-10$ (dashed curves) and $m=11-15$ (solid curves) on cylinder permittivity for size parameter $kR=10$. Vertical black arrows corresponds to $\varepsilon=3.4975$ (near TE$_{11,2}$ Mie resonance, red curve) and to $\varepsilon=3.3606$ (near TE$_{14,1}$ Mie resonance, magenta curve).}
\label{mie resonances}
\end{figure}

This fact is the prerequisite of the pronounced interference. As we have already seen, we may adjust this interference so that to obtain a deep and narrow minimum of the electric field $E_{\rho}=E_y= (i\omega\varepsilon_0\rho)^{-1}\partial H/\partial \phi$ on the axis $y$ behind the cylinder. In this minimum, the electric field of the $M$-th mode and that of the quasi-continuum have the same magnitudes and opposite phases. Inside the cylinder near its back edge their interference is constructive and $E(\phi=0,\rho)=E_{\rho}$ has a local maximum. A pair of maximum and minimum, adjacent to {one another over the axis $y$,} can be treated as the spatial Fano resonance. When a Fano resonance occurs in the frequency domain{,} 
the minimum neighboring to the Fano maximum is always deep and narrow. Is it possible to observe a similar sharp minimum in our spatial Fano resonance? Yes, {it needs an interplay between three parameters: the optical size of the cylinder $kR$, the permittivity of the cylinder $\varepsilon$ and the illumination angle of two plane waves $\beta$. We may find this regime either for any fixed $\varepsilon$ or fixed size parameter $kR$ and varying the other two parameters to get a sharp minimum. In this section we study this regime for fixed size parameter and in section \ref{optimization} we elaborate more on a practical approach to find this regime for fixed $\varepsilon$.} {Dependence of Mie coefficients on $\varepsilon$ for fixed size parameter $kR=10$ is presented in Fig. \ref{mie resonances}. Based on these coefficients we can calculate the intensity in the area of our interest.}

\begin{figure*}[t]
\centering
\includegraphics[width=15cm]{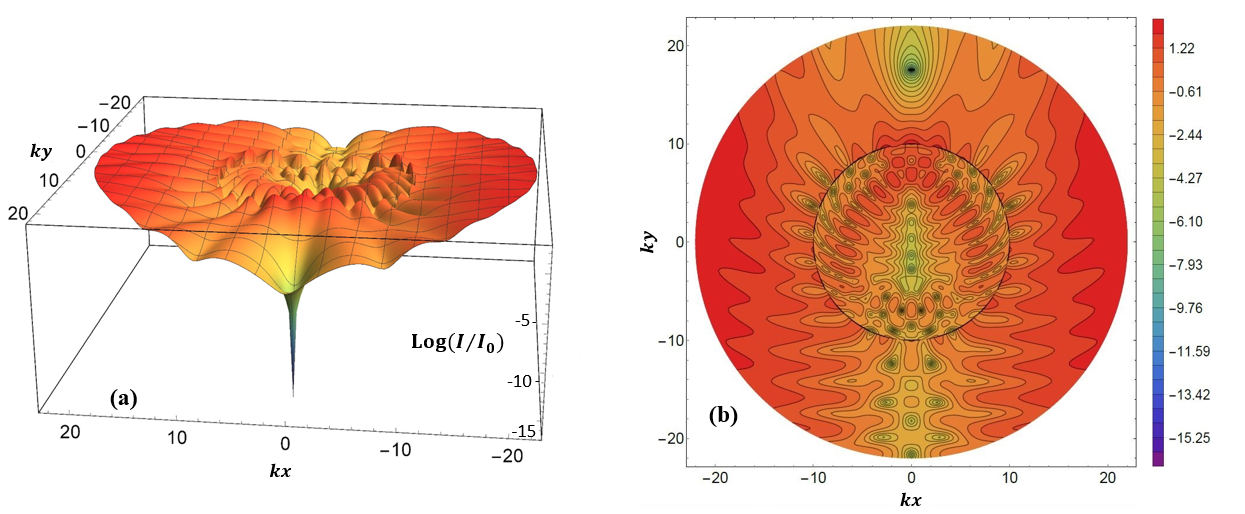}
\caption{{Normalized intensity $I/I_0$ in log scale in the cross section of the cylinder: (a) 3D plot  (b) 2D color map. Fixed parameters: $\beta=0.01$, $kR=10$, $\varepsilon=3.4975$ TE$_{11,2}$ Mie resonance. }}
\label{Pic2}
\end{figure*}

Since the incident beam intensity varies versus $x$, it is reasonable to normalize the intensity of the total electric field $I=E_{\rho}^2+E_{\phi}^2$ to the intensity of the incident beam $|E_{ib}|^2$, averaged over the relevant interval $-R<x<R$. The integration is  simplified by the condition 
$\beta\ll 1$, and we obtain {$I_0=(2\eta H_0)^2\sin^2(kR\sin\beta)/\sin^2\beta$.} 
On the axis $y$ the incident and total electric fields are polarized longitudinally and the total intensity is equal $I(\phi=0,\rho)=[E_{\rho}(\phi=0)]^2$. 
We normalize it to the mean intensity $I_0$ of the incident beam.  

The logarithmic plot of the normalized intensity {in the area of our interest is presented for some choices of $\varepsilon$ (from the values close to the resonant ones) in Fig.~\ref{Pic1}(b). The results are based on the choice of $\beta=0.01$ for illumination angle.} The cylinder permittivity $\varepsilon=2.4445$ (corresponds to TE$_{12,1}$ Mie resonance) offers a weak  minimum to the normalized intensity at the distance of the order of $\lambda$ behind the cylinder. 
{In Fig.~\ref{mie resonances} we can see how to tune the Fano resonance varying $\varepsilon$ near the region of Mie resonanes.} 
In this way we found several values of $\varepsilon$ corresponding to ultimately narrow and deep minima of the electric field behind the cylinder. Two of them are marked in Fig.~\ref{Pic1}(b) by vertical arrows. Permittivities $\varepsilon=2.908$ (corresponds to TE$_{13,1}$ Mie resonance) and $\varepsilon=3.4975$ (near TE$_{11,2}$ Mie resonance), offer intensity in  minimum smaller than $I_0$ by 10 and 14 orders of magnitude, respectively. Since the magnetic field on the axis $y$ is identically zero, in these minima the electromagnetic field practically vanishes.

Distribution of the normalized electric intensity in the plane $(x-y)$ confirms our insight that these minimas are namely those of spatial Fano resonances. In Fig.~\ref{Pic2}(a) we present the plot of $\log{(I/I_0)}$ in the plane $(x-y)$ for $\beta=0.01$ and $\varepsilon=3.4975$.   
Internal maxima corresponding to the almost-resonant mode TE$_{11,2}$ are located around the cylinder near its surface. One of these maxima is higher than the others and is located at the axis $y$. It forms together with our minimum a typical Fano resonance. Meanwhile, in the color map  Fig.~\ref{Pic2}(b) we see that the electric field distribution is very different from the typical picture of a mode $M\gg 1$ in the range of its resonance excited by a single plane wave \cite{Kavungal}. The distorted modal distribution with sharp interference minima is explained by the interference of the $M$-th almost resonant mode and the quasi-continuum of lower modes which have the same magnitudes in the region of our interest.

\begin{figure*}[htbp]
\centering
\includegraphics[width=15cm]{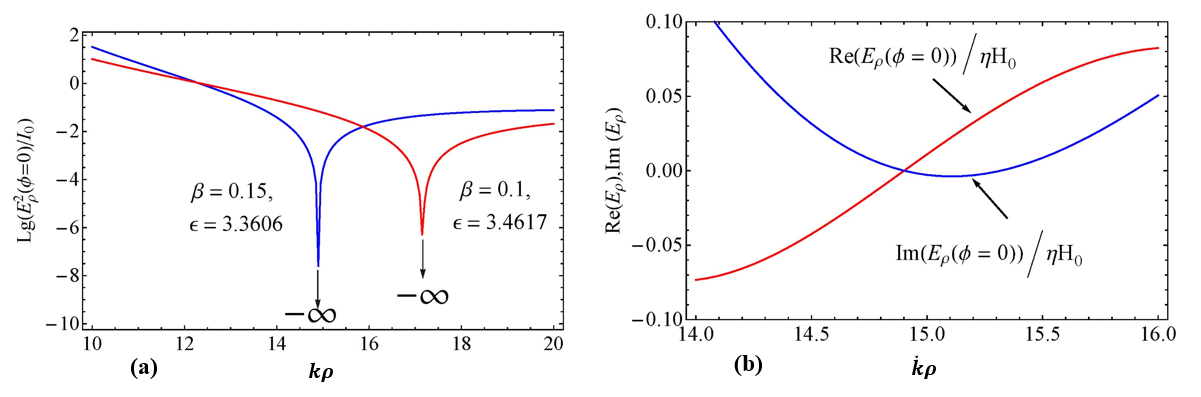}
\caption{(a) Normalized intensity plot in log scale for two pairs of $\beta$ and $\varepsilon$ offering the exact zero for the electromagnetic field. (b) Real and imaginary parts of the normalized electric field phasor in the area of the Fano minimum for $\beta=0.15$, 
$\varepsilon=3.360595$. Fixed parameter: $kR=10$. }
\label{Pic3}
\end{figure*}

Real and imaginary parts of the electric field phasor changes sign at different points of axis $y$, and our Fano minimum lies between these points being distant from the cylinder by $\rho-R\approx 0.7\lambda$. In the vicinity of our minimum the phase of the electric and magnetic fields differ by nearly $\pi/2$, that clearly links the effect to evanescent waves generated on the back surface of the cylinder. Conventionally, near-field effects cannot be observed far from the scattering object and they enhance the local fields. Our near-field effect is opposite -- it is cancellation of the small longitudinal component of a wave beam by the evanescent waves and it may nicely occur at distances about $\lambda$ from the object. 

\begin{figure*}[t]
\centering
\includegraphics[width=15cm]{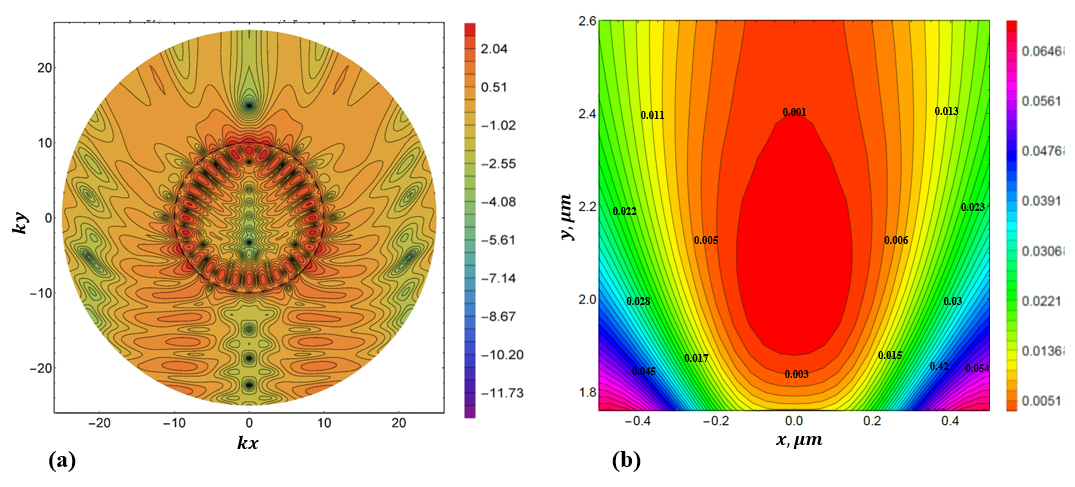}
\caption{(a) Color map of {$\log{(I/I_0)}$} in the cylinder cross section . 
(b) Color map of $I/I_0$ (linear scale) in the area of the optical trap. Fixed parameters: $kR=10$, $\beta=0.15$ and $\varepsilon=3.360595$.}
\label{Pic4}
\end{figure*}

\begin{figure*}[t]
\centering
\includegraphics[width=15cm]{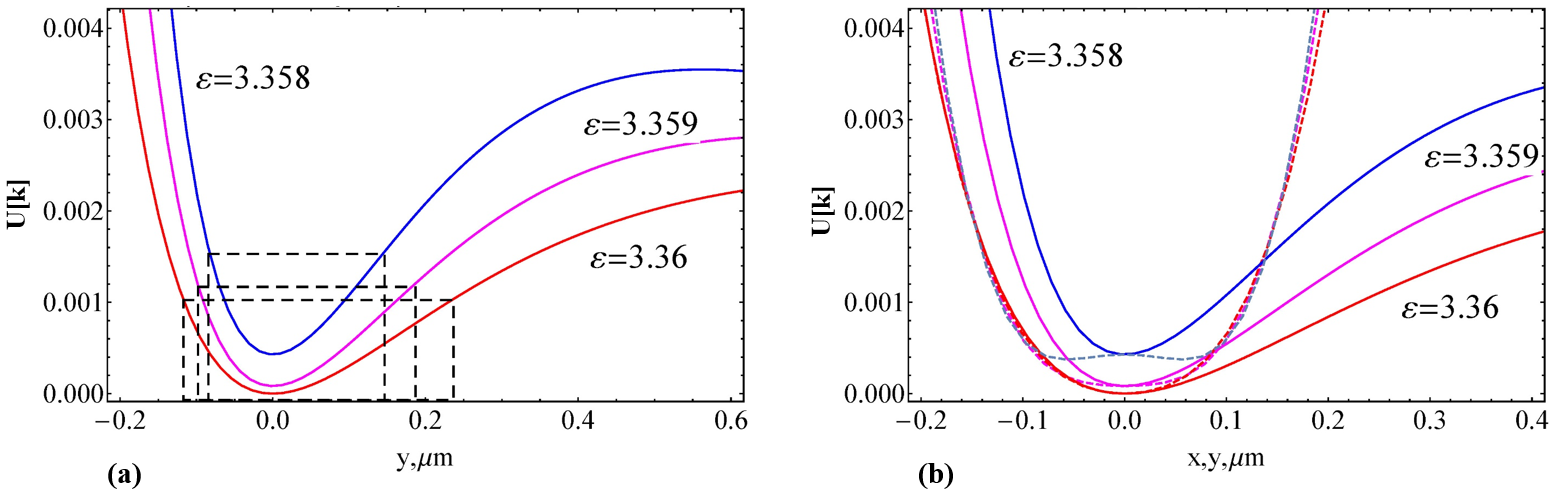}
\caption{(a) The optical potential of an atom of Cs varying along the beam axis for different dielectric constants of the cylinder material. The black dashed rectangles show the width of the trap that can be reduced to 230 nm adjusting $\varepsilon$. (b) The same potentials depicted over both $x$ and $y$ axes allow one to compare the trap width and height (both are subwavelength).
Transition in Cs is enabled by laser radiation at wavelength $\lambda=852 nm$. Poynting vector magnitude at maximum of the incident beam intensity is equal to $1$ kW/cm$^2$. Fixed parameters: $kR=10$ and $\beta=0.15$.}
\label{Pic5}
\end{figure*}

\begin{figure*}[t]
\centering
\includegraphics[width=17cm]{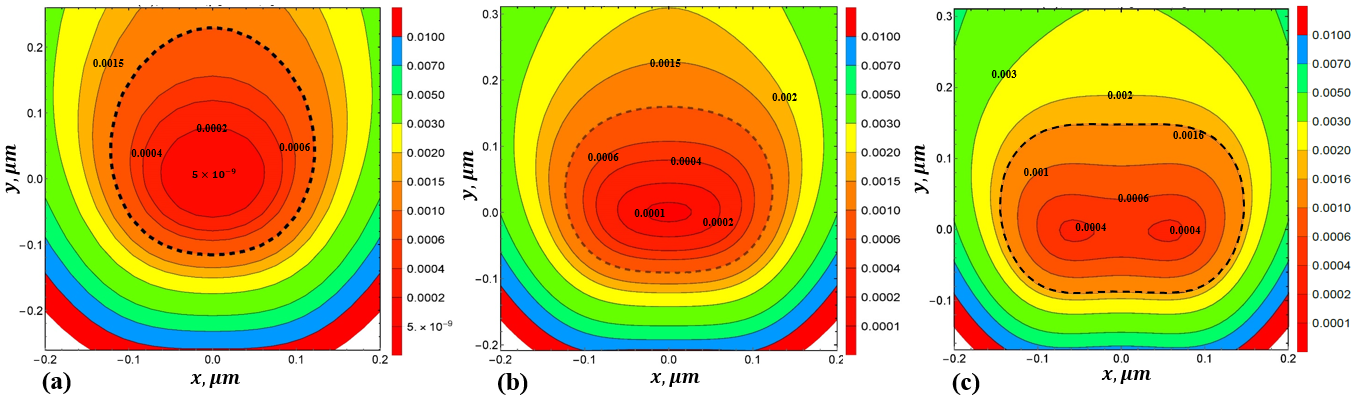}
\caption{Contour plots of the optical potential of an atom of Cs: (a) $\epsilon=3.36$, (b) $\epsilon=3.359$, and (c) $\epsilon=3.358$. Dashed contours
{depict levels of optical potential which are 1 mK higher than the local minimum}. It defines the effective perimeter of the trap. Fixed parameters: $kR=10$ and $\beta=0.15$.}
\label{Pic6}
\end{figure*}

\begin{table*}[t]

\caption{\label{tab}Main properties of alkaline atoms.}

\begin{tabular}{ | m {4em} | m{7em}| m{1.3cm} | m{2.3cm} | m{1.4cm} | m{3.3cm} |m{3.3cm} |} 
\hline
\hline

Atoms & Transition & $\lambda$ [nm] & Decay Rate $s^{-1}$ & Life time $nm$ & Transition Dipole
Matrix Element SI ($J=1/2||er||J'=3/2)$ & Transition Dipole Matrix Element SI ($J=1/2||er||J'=3/2)/\sqrt{2}$)\\ 
\hline

$^{133}Cs$ $D_2$ & $6^2S_{1/2}$→$6^2P_{3/2}$ & 852.347 & $3.281 \times10^7$ & 30.47 & $3.797 \times10^{-29}$ & $2.68 \times10^{-29}$ \\ 

$^{87}Rb$ $D_2$ & $5^2S_{1/2}$→$5^2P_{3/2}$ & 780.241 & $3.811 \times10^7$ & 26.24 & $3.584 \times10^{-29}$ & $2.53 \times10^{-29}$\\ 

$^{23}Na$ $D_2$ & $3^2S_{1/2}$→$3^2P_{3/2}$ & 589.158 & $6.154 \times10^7$ & 16.25 & $2.988 \times10^{-29}$ & $2.11 \times10^{-29}$ \\ 
\hline
\hline
\end{tabular}
%\end{center}

\end{table*}

{ For fixed size parameter (in our case $kR=10$), by tuning $\varepsilon$ and $\beta$ we can nullify ${\rm Re}(E_\rho)$ and ${\rm Im}(E_\rho)$ at the same point on the axis $y$. For two different illumination angle this tuning is made and the results in log scale of intensity are presented in Fig.~\ref{Pic3}(a). For case $\beta=0.15$ we found $\varepsilon=3.360595$ and zero point is distanced from the cylinder still by $0.7\lambda$ .For $\beta=0.1$ we found $\varepsilon=3.4617233453067$ granting the similar zero at the distance nearly equal $\lambda$. Thus, behind the cylinder there is an amazing point in which ${\rm Re}(E_\rho)={\rm Im}(E_\rho)=0$ and electromagnetic fields vanishes. This situation is shown in Fig.~\ref{Pic3}(b) for the case $\beta=0.15$, $\varepsilon=3.360595$.}

In Fig.~\ref{Pic4}(a) we depict the color map of the normalized intensity in the plane $(x-y)$  for the case $\beta=0.15$, $\varepsilon=3.360595$. This distribution resembles the picture of the TE$_{14,1}$ resonance excited by a single plane wave. However, there is a difference -- a set of sharp minima. Four of them are located on the axis $y$ outside the cylinder. {Only at the minimum located behind the cylinder, the exact zero is achieved.} Three minima located in front of it are much weaker -- in them $I/I_0\sim 10^{-3}-10^{-4}$. Fig.~\ref{Pic4}(b) represents the same color map as in Fig.~\ref{Pic4}(a) as a contour plot shown around the main Fano minimum. This plot lets one see that the shape of the minimum is not circular, it is elongated in the axial direction.   

\section{Optical trap at the Fano minimum}

The revealed effect is, to our opinion, very promising for trapping the atoms and ions. An atom with polarizability $\alpha$, experiences in the monochromatic light of non-uniform intensity $I(x,y)$ the so-called gradient force 
$$ F_g(x,y)={1\over 2}{\rm Re}(\alpha)\nabla I(x,y).$$ 

This formula, initially derived in \cite{Ashkin} for dielectric nanoparticles, was generalized for atoms in the laser light field in \cite{Bradshaw, handbook}. In case of blue detuning (from the main excited state of the atom) ${\rm Re}(\alpha)<0$ and direction of $F_g$ will be  toward the minimum of the electric intensity \cite{trap3}. Since our minima in Figs.~\ref{Pic2} and~\ref{Pic3} are ultimately sharp, {an atom will be trapped at the small region in the $(x-y)$ plane and free to move along $z$}. With a simple glass microcylinder we may prepare a unique object --  a straight linear chain of atoms in free space.  

Our singular Fano minima can serve as a trap for ions too. In accordance to \cite{Poles}, 
the paraxial region of a Bessel beam with radial polarization and nonzero order can serve a subwavelength thin trap for charged particles. It is possible to show that this trapping property {keeps} in our 2D case when the Bessel beam is replaced by our two-wave beam. {Gradient force will gather the ions to our intensity minimum from the surrounding space.}

When the incident light has the same wavelength as that of the resonant optical absorption $\lambda_A${,} or is slightly detuned (so that $\lambda$ is within the optical transition spectral line) and {has sufficiently} high flux density{,} the optical transition of an atom at $\lambda_A$ is pumped and the atom turns out to be cooled \cite{trap1}. When the flux density is of the order of one kW per square cm, {optical} potential of an atom expressed in Kelvins is of the order of dozens of mK. If the frequency detuning $\Delta=\omega-\omega_A$ of the incident light with respect to the transition frequency is positive, the polarizability of an atom has the negative real part and the trapping effect arises in the minimums of the electric intensity where the atom {optical} potential drops from dozens of mK to one mK or less \cite{trap1,trap2,trap3}. Usually the alkaline atoms are used for trapping. In table \ref{tab} one can see main properties of such atoms\cite{alkaline}

All these atoms have similar properties. As an example Figs~\ref{Pic5} and ~\ref{Pic6} depict the optical potential $U(x,y)$ in our optical trap for the case when the trapped atoms are those of Cs, $kR=10$ and $\beta=0.15$. 
They experience at the wavelength $\lambda_A=852$ nm the transition of type $6S_{1/2}-6P_{3/2}\, D2$. $U$ is calculated using the formula (see e.g. in \cite{trap2}) 
$$ 
U=\mu^2E^2/2k_B\hbar \Delta,
$$ 
where $\mu=2.68\cdot 10^{-29}$ SI is the dipole moment (matrix element) of the optical transition, $E^2=I$ is the electric intensity, corresponding to the power flux {$1$ kW/cm$^2$} in the intensity maximum of our beam, $k_B$ and $\hbar$ are Boltzmann and Planck constants, respectively. In these calculations, we adopted $\Delta=5\, GHz$. {The laser pumping with the flux density $1$ kW/cm$^2$ and even much higher one has been used for optical trapping since 1980s \cite{trap5}.} Approaching the minimum of the optical potential to zero offers the long-time confinement of atoms, because the atoms located at the minimum practically do not heat up.
The trap transverse sizes are sensitive to the design parameters. This effect is illustrated by Fig.~\ref{Pic5}(a) where the effective trap size in the $y$-direction
is shown for different values of $\varepsilon$. {A slight change in the permittivity (from 3.36 to 3.358) transforms the trap from that having the width of the order of 
$0.5\, \mu$m and the potential at the trap center close to 10 $\mu$K into a trap having twice smaller width but fifty times higher potential at the trap center.} Here the effective dimensions of the trap were estimated in accordance to the criteria formulated in \cite{trap4} ({1 mK above the potential minimal value}). {In Fig.~\ref{Pic5}(b) trap dimensions in both $x$ and $y$ directions are depicted. For better visualization, colour map of optical potential for the same parameters of Fig.~\ref{Pic5}(b) is shown in Fig.~\ref{Pic6} in which the dashed contours show the effective area of the trap. One can see from Fig.~\ref{Pic6} that the trap dimensions are subwavelength and dependent to value of absolute minimum of the potential in tarp region.} If we do not target the minimal possible size of the trap, {the absolute minimum of the potential in the trap region} can be engineered very close to zero because the exact zero of the electromagnetic field at the trap center is achievable.

\begin{figure}[h!]
\centering
\includegraphics[width=7cm]{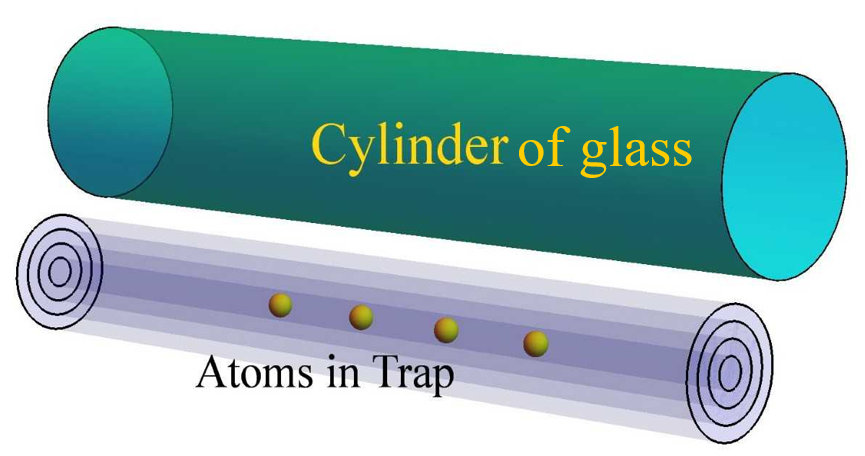}
\caption{Sketch of our optical trap.}
\label{Pic7}
\end{figure}

Thus, a simple dielectric microcylinder illuminated by an intensive cosine wave beam of coherent light with magnetic field polarized along the 
cylinder axis creates in free space a long optical trap with subwavelength cross section. This idea is illustrated by Fig.~\ref{Pic7}. 
Cold atoms or ions can be guided along this trap e.g. by a static electric field. Then the trap will operate as an atomic waveguide. 
The creation of such atomic waveguides is an actual problem of modern physics. Therefore, similar traps has been recently developed
and corresponding works forms a body of literature (see e.g. in \cite{trap6,trap7,trap8}). 
However, all these traps, to our knowledge, have been formed only inside the diffraction-free light beams. 
In our opinion, this approach to the creation of the cylindrical optical trap 
has an inherent drawback -- complexity of its implementation. 

No realistic wave beam is ideally diffraction-free. In focused Bessel beams of zero order
\cite{trap6}, where the atoms are trapped in the area of the maximal intensity  
centered by the beam axis, the effective length of the cylindrical trap (whose cross section has the diameter close to $\lambda$)
does not exceed a dozen of microns. There are other problems with the trapping of atoms in these light needles  
where are studied in work \cite{trap7}. In this paper it was shown that the trapping in the zero-order diffraction-free beam (that demands an expensive equipment)
has no practical advantages compared to the trapping in the usual Gaussian beam formed by a standard laser optics. 
Only a hollow Bessel beam with radial polarization which remains diffraction-free up to the distance of hundreds of microns from its birthplace
(the apex of an axicon lens) grants a really thin and long cylindrical atom trap similar to what we suggest. Such the trap was demonstrated in work \cite{trap8}.
However, in order to obtain such a magnificent light beam one needs a very expensive optical equipment, more expensive than is required for a zero-order Bessel beam. On the contrary, our wave beam is a simple superposition of two plane waves -- practically of two laser beams with flat phase fronts, e.g. of two Gaussian beams. 

One may compared our trap with that suggested in work \cite{trap11}. However, this simple trap not only has different underlying physics, it is formed 
on the surface of a cylinder. Our trap is formed in free space at the distance of the order of $\lambda$ from the cylinder.
In work \cite{PNJ1} it was suggested to trap nanoparticles and molecules in the waists of the twin photonic nanojet obtained in this work impinging a microcylinder by two plane waves. In some sense this work is close to ours, however, the underlying physics and the results of \cite{PNJ1} are very different. 
In this work the plane waves are TE-polarized with respect to the cylinder, and the Mie resonances of the TM-type are excited in it. 
This beam polarization allows the excitation of two nanojets (which are prohibited in our case) and makes our Fano resonance impossible. 
The waists of the two nanojets obtained in \cite{PNJ1} have the slightly subwavelength width but have the substantial length. Finally, in \cite{PNJ1} the trapping is 
implied in the two maxima of intensity. On the contrary, our trap can have both subwavelength height and width, and, what is even more important, 
our trapping occurs in the minimum of the field (where the heating of the atom is either weak or even practically absent). 

As to the realization of our trap, all we need is two collimated laser beams having the flat fronts in the area of our cylinder. For fine tuning the parameter 
$\beta$ we may split a collimated laser beam onto two beams with equivalent amplitudes using the conventional optical scheme -- a semitransparent mirror and a fully reflecting one. The tilt of the reflecting mirror will ensure the needed $\beta$. Perhaps, one will need a diaphragm to get rid of the side-lobes. Tunable laser sources will allow one to obtain the needed operation frequency if the permittivity of our cylinder is predefined. A microcylinder can be either a cylindrical optical resonator (rod resonator), used for laser applications and in optical sensing or even a piece of the optical fiber. 
{Usual precision of the fiber manufacturing is about $10^{-3}$ of the diameter. For a fiber with radius $R=1\, \mu$m it implies 2 nm roughness. Such imperfection of a cylinder cannot be important because} evanescent waves with very high spatial frequencies are not involved into our effect.

\begin{figure}[h!]
\centering
\includegraphics[width=8.5cm]{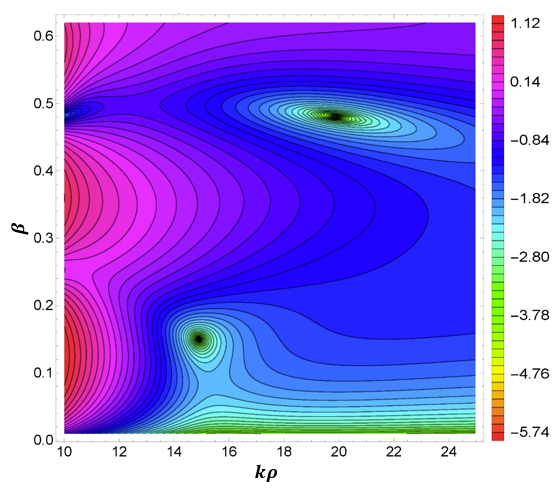}
\caption{The color map of normalized intensity (logarithmic scale) on the axis $y$ versus both normalized coordinate $ky=k\rho$ and $\beta$ for TE$_{14,1}$ resonance. Fixed parameters: $kR=10$ and $\varepsilon=3.3606$.}
\label{new3}
\end{figure}

\begin{figure}[h!]
\centering
\includegraphics[width=7.5cm]{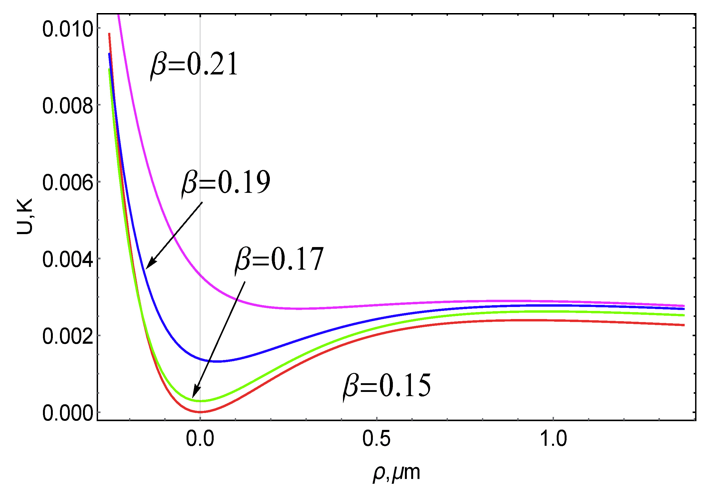}
\caption{The dependence of the trap potential {on the physical coordinate $y=\rho$ counted from the center of the trap }for different values of $\beta$ for TE$_{14,1}$ resonance. Fixed parameters: $kR=10$ and $\varepsilon=3.3606$.}
\label{new4}
\end{figure}

\section{Stability of the suggested trap}

\begin{figure*}[t]
\centering
\includegraphics[width=13.5cm,height=5.6cm]{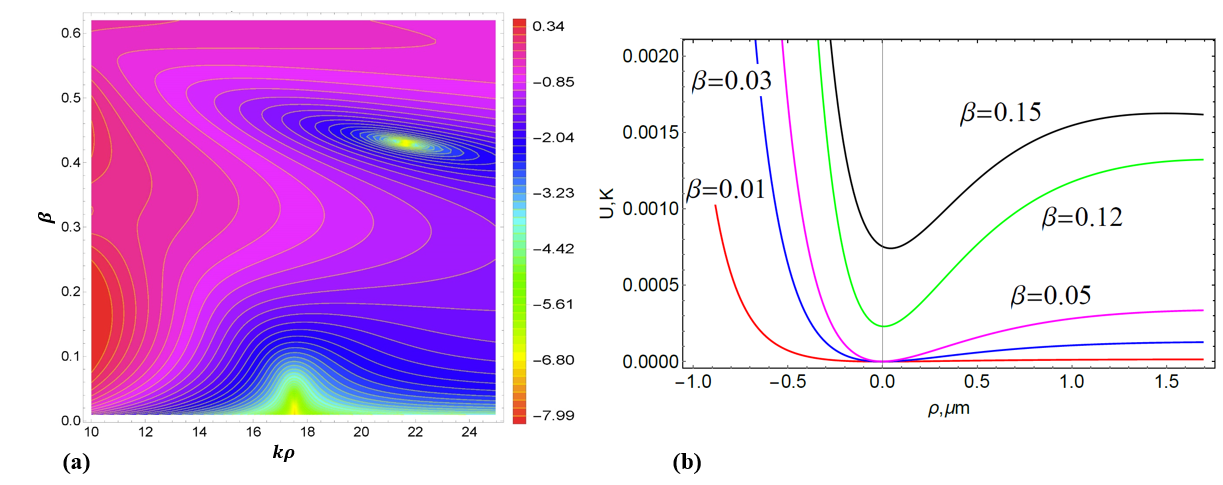}
\caption{{ The dependence of (a) normalized intensity (logarithmic scale) on $ky=k\rho$  and (b) the trap potential on the coordinate $y=\rho$ counted from the center of the trap, for different values of  $\beta$  for TE$_{11,2}$ resonance. Fixed parameters: $kR=10$ and $\varepsilon=3.4975$.}}
\label{new5}
\end{figure*}

\begin{figure}[ht]
\centering
\includegraphics[width=7.5cm,height=5cm]{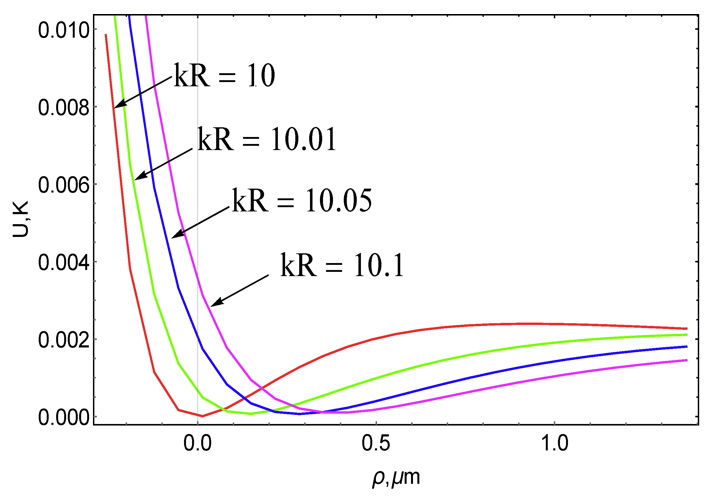}
\caption{ The dependence of the trap potential on the coordinate {$y=\rho$ counted from the center of the trap} for different size parameters 
close to $10$ when the TE$_{14,1}$ resonance is used. {Fixed parameters: $\beta=0.15$ and $\varepsilon=3.3606$.}}
\label{new6}
\end{figure}

\begin{figure}[h!]
\centering
\includegraphics[width=7.5cm]{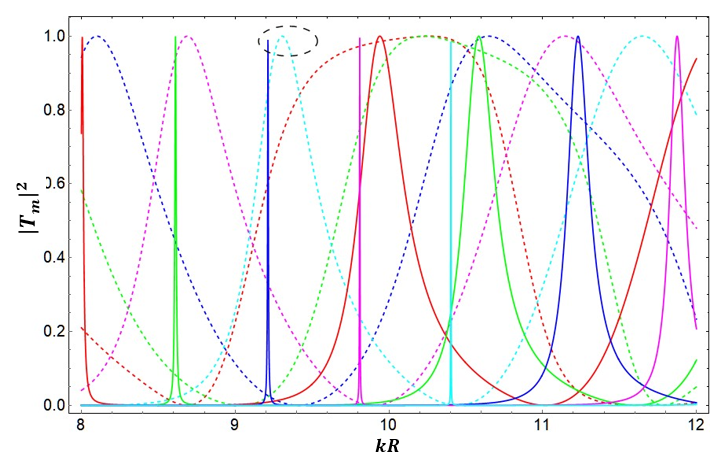}
\caption{The dependence of Mie coefficients on the size parameter for $\varepsilon=3.5$.}
\label{new7}
\end{figure}

\begin{figure}[t]
\centering
\includegraphics[width=7.9cm,height=6.0cm]{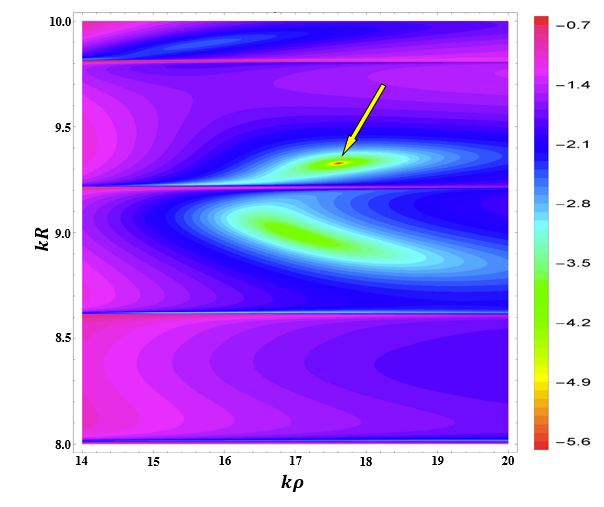}
\caption{ Distribution of the electric field intensity (logarithmic scale) along $y$ axis for different $kR$. Yellow arrow shows the first approximation for the trap optimal configuration. Fixed parameters: $\beta=0.15$ and $\varepsilon=3.5$.}
\label{new8}
\end{figure}

\begin{figure*}[ht]
\centering
\includegraphics[width=12.5cm,height=5.8cm]{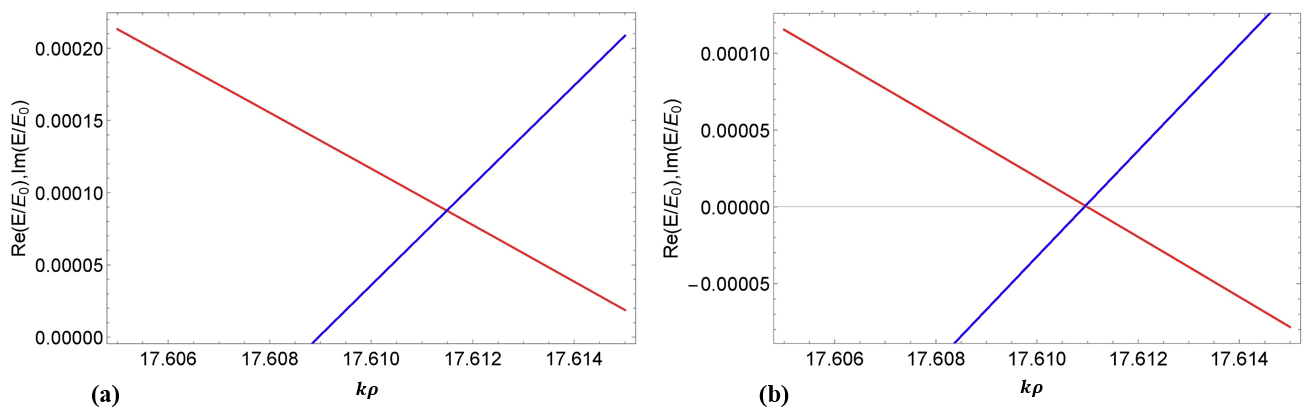}
\caption{ Real (red) and imaginary (blue) parts of $E_\rho(\phi=0,\rho)$ as a function of $k\rho$. (a) Initial distribution with arbitrary chosen $\beta=0.15$. 
(b) Optimized distribution achieved with $\beta=0.149914$. Fixed parameters: $kR=9.3278$ and $\varepsilon=3.5$. }
\label{new9}
\end{figure*}

{Our trap is sensitive to the incident beam parameters.} In Fig.~\ref{new3} the logarithm of normalized intensity on the axis $y$ behind the cylinder is shown dependent on both coordinate and $\beta$ for the case $\varepsilon=3.3606$. In Fig.~\ref{new4} the trap potential of the same system for different values of $\beta$ is plotted.  From Figs. \ref{new3} and \ref{new4} we see that the subwavelength trap disappears with the 10\% deviation of the incident beam angle parameter from its optimal value $\beta=0.15$. Another interesting feature seen in Fig. \ref{new3} is the presence of a similarly deep trap for $\beta\approx 0.5$ for which $sin(m\beta) \sim 1$ (where $m=14$ is the number of the resonant mode) and the resonant mode again dominates over the quasi-continuum.

\begin{figure}[ht]
\centering
\includegraphics[width=7.5cm]{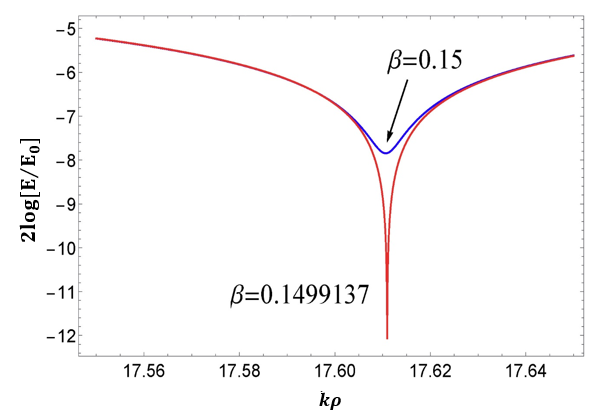}
\caption{Spatial distributions of the electrical field magnitude across the trap (log scale) for initially selected $\beta=0.15$ and for optimized $\beta=0.149914$. Fixed parameters: $kR=9.3278$ and $\varepsilon=3.5$.}
\label{new10}
\end{figure}

However, high sensitivity of the trap regime to $\beta$ is only a peculiarity of the selected Mie resonance. If we consider e.g. a trap which is formed when $\varepsilon=3.4975$ near the resonance TE$_{11,2}$  we see {that the trap is more stable} against small deviations of the angle $\beta$. In Fig. \ref{new5}(a) we presented a color map of the normalized intensity similar to that depicted in  Fig. \ref{new3} but for $\varepsilon=3.4975$. {From Fig. \ref{new5}(b) we can see that the trap is formed in a broad range of $\beta$. The exact null at the trap center is achieved  for $\beta=0.03$, and keeps negligibly small in the range of $0<\beta<0.05$.} The regime  without the exact zero of the field, for example, that corresponding to  $\beta=0.12-0.15$ grants a narrower trap. 
However, the potential (and the field) at its center become noticeable in this case. It means some heating {of} the trapped atoms that reduces their trapping lifetime.  There is a {trade off} between the maximal trapping lifetime and the minimal trap  width. For long trapping  small values of $\beta$ are preferable, however, the trap is rather wide (few microns). {For rather smaller cross section of the trap,} values $\beta=0.12-0.15$ are preferable but the trapping lifetime of an atom will {reduce} compared to that achievable for a more substantial trap.

Now, let us study the sensitivity of our trap to the variation of $kR$. {In practice, the cylinder radius $R$ and the permittivity $\varepsilon$ 
are fixed. We can tune our trap changing $\beta$ and $kR$. The size parameter is turned by varying the frequency. In Fig. \ref{new6} one can see how this frequency tuning affects the potential distribution on the axis $y$ for $\beta=0.15$ and $\varepsilon=3.361$. Varying $kR$ we not only displace the center of our trap, but also change the trap dimensions and trapping lifetime.}

\section{Practical approach to optimize the trap configuration }
\label{optimization}

{So far we have analysed the effect of a deep subwavelength trap offered by a microcylinder illuminated by two plane waves. In our examples, the permittivity values were adjusted. In this section we suggest a systematic approach to find the trap configuration for arbitrary experimental condition.} To exactly nullify the electric field intensity at a point ($\phi=0$, $\rho=\rho_c>R$) we have to satisfy two equations:
$$
{\rm Re}E_{\rho}(k\rho_c,\beta,kR,\varepsilon)=0, \quad {\rm Im}E_{\rho}(k\rho_c,\beta,kR,\varepsilon)=0. 
$$
For fixed $\varepsilon$ these two equations define a line $\rho_c(\beta,kR,\varepsilon)$, on which any point corresponds to a zero-field trap. 
Using the expression (4) for the electric field we have found several lines $\rho_c(\beta,kR,\varepsilon)$ near the high-order Mie resonances.
Their existence proves the variety of such zero-field traps. However, this direct trap engineering is not an easy numerical task since the calculation of the
line $\rho_c(\beta,kR,\varepsilon)$ implies the solution of an inverse problem.
Therefore, we have also developed a rather straightforward optimization procedure.  
In accordance to this simple procedure, one should implement the following steps:
\begin{enumerate}
 \item Choose specific material and its specific permittivity $\varepsilon$. 
 \item Choose resonant size parameter $kR$ of the cylinder. It can be found by analyzing Mie coefficients $T_m$. In experiment it can be performed by frequency tuning.
 \item Finely tune size parameter $kR$ and beam angle $\beta$ to get deep trap. In experiment it can be performed by frequency and optical elements tuning.
   \end{enumerate}

{This procedure is illustrated by an example of a popular material with $\varepsilon=3.5$ (so-called OHARA LAH size 75 optical glass). To choose the resonance region 
we plot the dependence of Mie coefficients on the size parameter $kR$ for $\varepsilon=3.5$ Fig. \ref{new7}. On this plot we select a suitable resonance, e.g. that marked on the plot corresponding to TE$_{13,1}$ and TE$_{10,2}$ modes. Then choose and initial value for $\beta$, here for example $\beta=0.15$, and find value of $kR$ which provides the deepest trap. In our case we found $kR=9.3278$ as the first approximation of our trap configuration, as is pointed out via yellow arrow in Fig. \ref{new8}. After trap position is found within first approximation, next step is to finely tune $\beta$ by considering real and imaginary parts of $E_\rho$. The target is to nullify the field in the vicinity of this point. The result of this numerical tuning of $\beta$ is presented in Fig. \ref{new9}. Fig. \ref{new10} shows very deep trap configuration for OHARA LAH75 which nicely grants a very deep and narrow trap.}

\section{Conclusions}

In this work we have theoretically studied an unusual near-field effect -- spatial Fano resonance arising when the wave beam formed by two plane waves 
with the small angle between their wave vectors impinges a dielectric microcylinder. {The physics of this phenomenon is the interference of 
the evanescent field of a nearly-resonant cavity mode with the quasi-continuum of all other modes.} We have shown that the Fano minimum is located at a substantial distance from the rear edge of the microcylinder, this is very unusual for a near-field effect. Our main result is the singularity of this minimum -- the electromagnetic field in it decreases very sharply and may even {utterly nullify.} Practically{,} it results in the possibility of creation a long optical trap that seems to be promising for cold atoms and ions. Such a trap can be called a particle waveguide. The evident advantage of our optical trap compared to its known analogues is the simplicity of its implementation. We hope that our finding will be interesting for physicists developing optical traps, especially particle waveguides for quantum computing \cite{Quantum1, Quantum2}.

\section{Acknowledgment}

Funding by Russian Foundation for Basic Research (grant number: 18-02-00315) is acknowledged by V.K.

\bibliography{PRB.bib}
\end{document}